\documentclass[twocolumn]{aastex631}
\usepackage{amsmath}

\begin{document}

\newcommand{\pfh}[1]{\textcolor{red}{#1}}

\newcommand{\GIZMO}{{\small GIZMO }}
\newcommand{\SF}{{\small STARFORGE }}

\newcommand{\gizmourl}{\href{http://www.tapir.caltech.edu/~phopkins/Site/GIZMO.html}{\url{http://www.tapir.caltech.edu/~phopkins/Site/GIZMO.html}}}
\newcommand{\datastatement}[1]{\begin{small}\section*{Data Availability Statement}\end{small}{\noindent #1}\vspace{5pt}}
\newcommand{\microGauss}{\mu{\rm G}}
\newcommand{\Bangle}{\theta_{B}}
\newcommand{\Alf}{{Alfv\'en}}
\newcommand{\BV}{Brunt-V\"ais\"al\"a}
\newcommand{\fref}[1]{Fig.~\ref{#1}}
\newcommand{\sref}[1]{\S~\ref{#1}}
\newcommand{\aref}[1]{App.~\ref{#1}}
\newcommand{\tref}[1]{Table~\ref{#1}}

\newcommand{\Dt}[1]{\frac{\mathrm{d} #1}{\mathrm{dt}}}
\newcommand{\initvalupper}[1]{#1^{0}}
\newcommand{\initvallower}[1]{#1_{0}}
\newcommand{\driftvel}{{\bf w}_{s}}
\newcommand{\driftvelmag}{w_{s}}
\newcommand{\driftvelhat}{\hat{{\bf w}}_{s}}
\newcommand{\driftveli}[1]{{\bf w}_{s,\,#1}}
\newcommand{\driftvelmagi}[1]{w_{s,\,#1}}
\newcommand{\dustvel}{{\bf v}_{d}}
\newcommand{\gasvel}{{\bf u}_{g}}
\newcommand{\gasden}{\rho_{g}}
\newcommand{\rhobase}{\rho_{\rm base}}
\newcommand{\gaspressure}{P}
\newcommand{\dustden}{\rho_{d}}
\newcommand{\rhodust}{\dustden}
\newcommand{\rhogas}{\gasden}
\newcommand{\resolution}{\Delta x_{0}}
\newcommand{\opticaldepth}{\tau_{\rm ext}}
\newcommand{\tauparam}{\tau_{\rm SL}}
\newcommand{\ts}{t_{s}}
\newcommand{\cs}{c_{s}}
\newcommand{\vA}{v_{A}}
\newcommand{\tL}{t_{L}}
\newcommand{\grainsuff}{_{\rm grain}}
\newcommand{\internaldensity}{\bar{\rho}\grainsuff^{\,i}}
\newcommand{\grainsize}{\epsilon\grainsuff}
\newcommand{\grainsizebar}{\bar{\epsilon}\grainsuff}
\newcommand{\grainmass}{m\grainsuff}
\newcommand{\graincharge}{q\grainsuff}
\newcommand{\grainchargeZ}{Z\grainsuff}
\newcommand{\grainsizemax}{\grainsize^{\rm max}}
\newcommand{\grainsizemin}{\grainsize^{\rm min}}
\newcommand{\B}{{\bf B}}
\newcommand{\Bmag}{|\B|}
\newcommand{\Bhat}{\hat\B}
\newcommand{\bhat}{\Bhat}
\newcommand{\acc}{{\bf a}}
\newcommand{\Lbox}{L_{\rm box}}
\newcommand{\Lscale}{H_{\rm gas}}
\newcommand{\sizeparam}{\tilde{\alpha}}
\newcommand{\sizeparammax}{\sizeparam_{\rm m}}
\newcommand{\chargeparam}{\tilde{\phi}}
\newcommand{\chargeparammax}{\chargeparam_{\rm m}}
\newcommand{\accparam}{\tilde{a}_{\rm d}}
\newcommand{\accparammax}{\tilde{a}_{\rm d,m}}
\newcommand{\accabsmax}{{a}_{\rm d,m}}
\newcommand{\accsizedep}{\psi_{a}}
\newcommand{\gravparam}{\tilde{g}}
\newcommand{\dustgas}{\mu^{\rm dg}}
\newcommand{\dustgashat}{\hat{\mu}^{\rm dg}}
\newcommand{\angstrom}{\mbox{\normalfont\AA}}

\def\app#1#2{%
  \mathrel{%
    \setbox0=\hbox{$#1\sim$}%
    \setbox2=\hbox{%
      \rlap{\hbox{$#1\propto$}}%
      \lower1.1\ht0\box0%
    }%
    \raise0.25\ht2\box2%
  }%
}
\def\approxprop{\mathpalette\app\relax}

\newcommand{\nadine}[1]{\textcolor{red}{#1}}

\title{Are Stars Really Ingesting their Planets? Examining an Alternative Explanation}
\shorttitle{Stellar Pollution: Dust or Planets?}
\correspondingauthor{Nadine H.~Soliman}
\email{nsoliman@caltech.edu}
\author[0000-0002-6810-1110]{Nadine H.~Soliman}
\affiliation{TAPIR, Mailcode 350-17, California Institute of Technology, Pasadena, CA 91125, USA}
\author[0000-0003-3729-1684]{Philip F.~Hopkins}
\affiliation{TAPIR, Mailcode 350-17, California Institute of Technology, Pasadena, CA 91125, USA}

\begin{abstract}
Numerous stars exhibit surprisingly large variations in their refractory element abundances, often interpreted as signatures of planetary ingestion events. In this study, we propose that differences in the dust-to-gas ratio near stars during their formation can produce similar observational signals. We investigate this hypothesis using a suite of radiation-dust-magnetohydrodynamic {\small{STARFORGE}} simulations of star formation. Our results show that the distribution of refractory abundance variations ($\rm \Delta [X/H]$) has extended tails, with about 10-30\% of all stars displaying variations around $\sim$0.1 dex. These variations are comparable to the accretion of $2-5 \rm M_\oplus$ of planetary material into the convective zones of Sun-like stars. The width of the distributions increases with the incorporation of more detailed dust physics, such as radiation pressure and back-reaction forces, as well as with larger dust grain sizes and finer resolutions. Furthermore, our simulations reveal no correlation between $\rm \Delta [X/H]$ and stellar separations, suggesting that dust-to-gas fluctuations likely occur on scales smaller than those of wide binaries. These findings highlight the importance of considering dust dynamics as a potential source of the observed chemical enrichment in stars.

\end{abstract}



\section{Introduction} \label{sec:intro}

Recent high-precision spectroscopic studies challenge the assumption of chemical homogeneity in co-natal stellar populations, revealing substantial variations in refractory element abundances among stars in open clusters and wide binaries \citep[e.g.][]{desidera2006abundance, desidera2004abundance, liu2016hyades,liu2019chemical, hawkins2020identical, spina2018chemical, nagar2019chemical, spina2021chemical, liu2024least}. Notably, 10\% to 30\% of stars exhibit strong ($> 2\sigma$) variations \citep{spina2021chemical, liu2024least}, and around 20\% of wide binaries show significant ($>$0.08 dex) iron abundance differences \citep{hawkins2020identical}. Deviations of $>0.05$ dex are also noted in individual binary systems \citep[e.g.][]{ramirez2011, tucci2014, mack2014,teske2015,ramirez2015dissimilar,biazzo2015gaps, teske2016curious, saffe2016, melendez2017solar, liu2018detailed, oh2018kronos, maia2019revisiting, ramirez2019chemical, church2020super, galarza2021evidence}.

These anomalies occur among stars with similar atmospheric parameters, suggesting they cannot be explained by atomic diffusion or analysis systematics alone \citep{dotter2017influence}. This has led to alternative explanations, such as planetary engulfment, which requires a few Earth masses of material for enrichment \citep{pinsonneault2001mass, church2020super, liu2024least}. Planetary engulfment is most effective after the star reaches the main sequence, when its thin, stable convective envelope allows significant atmospheric enrichment \citep{laughlin1997possible}. In contrast, earlier ingestion would likely be diluted within the star’s thick outer layer \citep{spina2018chemical}. However, \citet{saffe2024disentangling} reported a $\rm \Delta [Fe/H] \sim 0.08$ between components of a giant-giant binary, suggesting that significant deviations can occur even in stars with thick convective layers. This implies either several Jupiter masses of refractory material are needed or an alternative mechanism, which \citet{saffe2024disentangling} prefer.

The planet ingestion hypothesis, though plausible, is not the only explanation for such chemical signatures. Another possible explanation involves ``primordial'' fluctuations in refractory-rich dust grains near the star during its early formation. Variations in the dust-to-gas ratio in the material accreted by the star could lead to chemical inhomogeneity compared to stars formed in regions with different ratios. Thus, the primary distinction is whether compositional deviations were imprinted onto the star early through dust grain accretion or later through planetary material ingestion. Both scenarios produce similar observational signals: variations in surface abundances of refractory elements, correlated with condensation temperatures. The similarity in trend with condensation temperature stems from the common process of elements depleting onto solids as they condense out of the gas phase \citep{yin2005dust, gaidos2015little}.

The primordial fluctuation mechanism differs from planetary ingestion in that it can cause both reductions and enhancements in surface abundances. Observations often cannot determine if the differences arise due to enhancements or reductions. However, the correlation between surface element deviations and condensation temperature can help distinguish the cause, where a positive correlation usually indicates enhancement, while a negative slope suggests reduction. Negative slopes have been reported \citep{gonzalez2010parent, ramirez2014chemical, adibekyan2016abundance}, but they might also result from material locked in terrestrial planets that was not accreted by the star \citep{melendez2009peculiar, ramirez2010possible}, and not necessarily primordial dust-gas fluctuations.

Previous studies have explored the role of preferential dust or gas accretion in driving chemical variations in stars.
For instance, \citet{spina2021chemical} examined this mechanism but found it unlikely to be significant. However, their analysis was confined to dust accretion from the protostellar disk and did not consider the entire mass accretion period.  \citet{gaidos2015little} discussed a more relaxed general version by considering the substantial fluctuations in dust-to-gas ratios that could arise in star-forming regions and concluded that these fluctuations could be a viable mechanism for driving chemical deviations.

It is well established that significant fluctuations in the dust-to-gas ratio occur on the scale of protostellar cores, both theoretically and observationally. Dust, behaving as charged aerodynamic particles, can decouple from gas dynamics, causing density fluctuations independent of the gas. These fluctuations are further intensified by external turbulence and instabilities \citep{hopkins_2014_totallymetal, moseley:2018.acoustic.rdi.sims, squire2018resonant, hopkins2022dust}. Observations have documented variations of $\sim 2-5$ orders of magnitude over scales from $\sim1$ pc to $\sim0.001$ pc \citep{thoraval:1997.sub.0pt04pc.no.cloud.extinction.fluct.but.are.on.larger.scales, thoraval:1999.small.scale.dust.to.gas.density.fluctuations, miville-deschenes:2002.large.fluct.in.small.grain.abundances, abergel:2002.size.segregation.effects.seen.in.orion.small.dust.abundances, flagey:2009.taurus.large.small.to.large.dust.abundance.variations, boogert2013infrared, pineda:2010.taurus.large.extinction.variations, alatalo:2011.ngc1266.molecular.outflow, pellegrini:2013.ngc.1266.shocked.molecules, nyland:2013.radio.core.ngc1266}. Numerical simulations confirm these findings, showing pronounced fluctuations in the dust-to-gas ratio under various conditions of star formation \citep{elperin:1996:grain.clustering.instability,  bracco:1999.keplerian.largescale.grain.density.sims, cuzzi:2001.grain.concentration.chondrules, johansen:2007.streaming.instab.sims, youdin:2007.turbulent.grain.stirring, youdin:2011.grain.secular.instabilities.in.turb.disks, carballido:2008.grain.streaming.instab.sims, bai:2010.grain.streaming.sims.test, bai:2010.streaming.instability, bai:2010.grain.streaming.vs.diskparams, pan:2011.grain.clustering.midstokes.sims, pan:2013.grain.relative.velocity.calc, bai:2012.mri.saturation.turbulence}.

Understanding the pathways that cause chemical variations is crucial for several reasons. First, it helps determine if these variations are due to planetary engulfment events, which could reveal how planetary systems evolve and the frequency of such dynamical changes, impacting our understanding of planetary orbit stability. Second, it is essential for evaluating the chemical homogeneity of stellar associations and the effectiveness of ``chemical tagging". Significant abundance variations within stellar groups could undermine chemical tagging as a tool for tracing a star’s progenitor cloud \citep{ness2018galactic}, making it important to assess the limits of this method.

To investigate whether fluctuations in the dust-to-gas ratio can lead to significant variations in refractory element abundances, we analyze a set of radiation-dust-magnetohydrodynamic (RDMHD) simulations of star-forming molecular clouds that explicitly incorporate dust dynamics (as described in \citet{soliman2024dust, soliman2024thermodynamics}. The simulations utilize the STAR FORmation in Gaseous Environments ({\small{STARFORGE}}) framework \citep{grudic2021starforge}, which offers comprehensive modeling of individual star formation—from protostellar collapse through subsequent accretion and stellar feedback to main-sequence evolution and stellar dynamics \citep{starforge.fullphysics}.

This letter is structured as follows: Section \ref{sec:2} provides an overview of the code and details the initial conditions (ICs) used in our simulations. In Section \ref{sec:res}, we present the results from our primary simulations and explore their implications, contrasting them with simulations that utilize simplified dust physics and varying grain sizes. Finally, Section \ref{sec:conc} summarizes our conclusions.

\section{Simulations}
\label{sec:2}
\subsection{\SF\,  Physics}

We analyze a set of STARFORGE simulations presented and detailed in \citep{soliman2024dust, soliman2024thermodynamics}, employing the \GIZMO code \cite{hopkins:gizmo} to simulate star-forming Giant Molecular Clouds (GMCs). These simulations utilize the \GIZMO Meshless Finite Mass MHD solver  \citep{hopkins:gizmo.mhd, hopkins:gizmo.mhd.cg} for solving the ideal magnetohydrodynamics equations, alongside the meshless frequency-integrated M1 solver for time-dependent radiative transfer equations \citep{lupi:2017.gizmo.galaxy.form.methods, lupi:2018.h2.sfr.rhd.gizmo.methods, hopkins:radiation.methods, grudic:starforge.methods, hopkins.grudic:2018.rp}. These simulations have been extensively compared to observational data, including properties of GMCs and statistics of the Initial Mass Function (IMF) \citep{guszejnov2021starforge, guszejnov2022effects, grudic2023does, millstone2023coevolution, hopkins2024forge}, studies of stellar dynamics such as multiplicity \citep{guszejnov2023effects}.

The simulations include sink particles representing individual stars, evolving through accretion and interacting with the medium via protostellar jets, winds, radiation, and, if criteria are met, supernovae \citep{starforge.fullphysics}. Radiation is discretized into five frequency bands coupled directly with the dust spatial distribution, while gas and dust undergoes cooling and heating as detailed in \citet{fire3}.

Individual stars are resolved and represented as sink particles, evolving through accretion of dust and gas and interacting with their environment via protostellar jets, winds, radiation, and potentially supernovae \citep{starforge.fullphysics, soliman2024dust}. Radiation is discretized into multiple frequency bands, directly influencing the dust distribution, while both gas and dust undergo cooling and heating processes as described in in \citet{fire3}.

\subsection{Dust Physics}

Dust grains are modeled as``super-particles'', similar to methods used in circumstellar disk simulations in \small{ATHENA}/\small{ATHENA++} \citep{bai2010particle, sun2023magnetohydrodynamic}. Each super-particle represents a population of grains with identical attributes such as size, mass, and charge, determined self-consistently by computing collisional, photoelectric, and cosmic ray charging rates \citep{draine:1987.grain.charging, tielens:2005.book}. We explicitly track the mass of accreted dust and gas over each sink particle's accretion history. Dust dynamics, including drag, Lorentz, gravity, and radiation pressure forces, are explicitly modeled. Individual grain trajectories are integrated, and local dust properties are interpolated to neighboring gas cells.

The grain sizes are statistically sampled to conform to a Mathis, Rumpl, and Nordsieck (MRN) size distribution \citep{mathis:1977.grain.sizes}, ensuring both the desired initial dust-to-gas ratio and a uniform particle distribution across logarithmic grain size intervals. Each grain retains its size and composition throughout the simulations, with no processes simulated that would create, destroy, or alter individual grain sizes. However, the local mean grain size could fluctuate as grains of different sizes, with varying dynamics, move in and out of regions.


\subsection{Initial conditions} 
\label{sec:initial}
As is standard for star formation simulations, the simulations are initialized as a uniform-density turbulent molecular cloud within a periodic box, surrounded by a diffuse warm ambient medium. The initial velocity distribution follows a Gaussian random field, with an initial virial parameter $\alpha_{\text{turb}} = 2$. The initial magnetic field is uniform, with a mass-to-flux ratio 4.2 times the critical value within the cloud.

We study two cloud configurations:

\begin{enumerate}
    \item A smaller cloud with $M_{\text{cloud}} = 2 \times 10^3 \rm M_{\odot}$ and radius $R = 3 \text{ pc}$, with a mass resolution of $\Delta m_{\rm gas} \sim 10^{-3} \rm M_{\odot}$.
    \item A larger cloud with $M_{\text{cloud}} = 2 \times 10^4 \rm M_{\odot}$ and radius $R = 10 \text{ pc}$, with a resolution of $\Delta m_{\rm gas} \sim 10^{-2} \rm M_{\odot}$.
\end{enumerate}

Dust super-particles have a mass resolution four times higher than the gas \citep{moseley:2018.acoustic.rdi.sims}. Due to the Lagrangian nature of our simulations, spatial resolution is adaptive, whereas mass resolution is more rigorously defined. Typically, in dense star-forming regions, resolutions are approximately ~10 AU for $\Delta m_{\rm gas} \sim 10^{-3} \rm M_{\odot}$ and ~100 AU for $\Delta m_{\rm gas} \sim 10^{-2} \rm M_{\odot}$.

The initial dust distribution maintains a statistically uniform dust-to-gas (DTG) ratio, $\rho_{\text{dust}} = 0.01 \rho_{\text{gas}}$, with $\rho_{\text{gas}}$ representing the gas density. Dust is composed of the standard MRN mixture of silicate ($\sim$60\%) and carbonaceous ($\sim $40\%) grains with a sublimation temperature around $1500 \text{ K}$ and an internal density of $\sim 2.25 \text{ g/cm}^3$. Grain sizes sample from the empirical power law model proposed by \citet{mathis:1977.grain.sizes}, characterized by $\frac{dn_{\text{d}}}{d\grainsize} \propto \grainsize^{-3.5}$. We consider maximum grain sizes of $0.1 \mu \text{m}$, $1 \mu \text{m}$, and $10 \mu \text{m}$, spanning a dynamic range where the maximum grain size is 100 times larger than the minimum size.

\section{Results}
\label{sec:res}
\begin{figure*}
    \centering
    \includegraphics[width=1\linewidth]{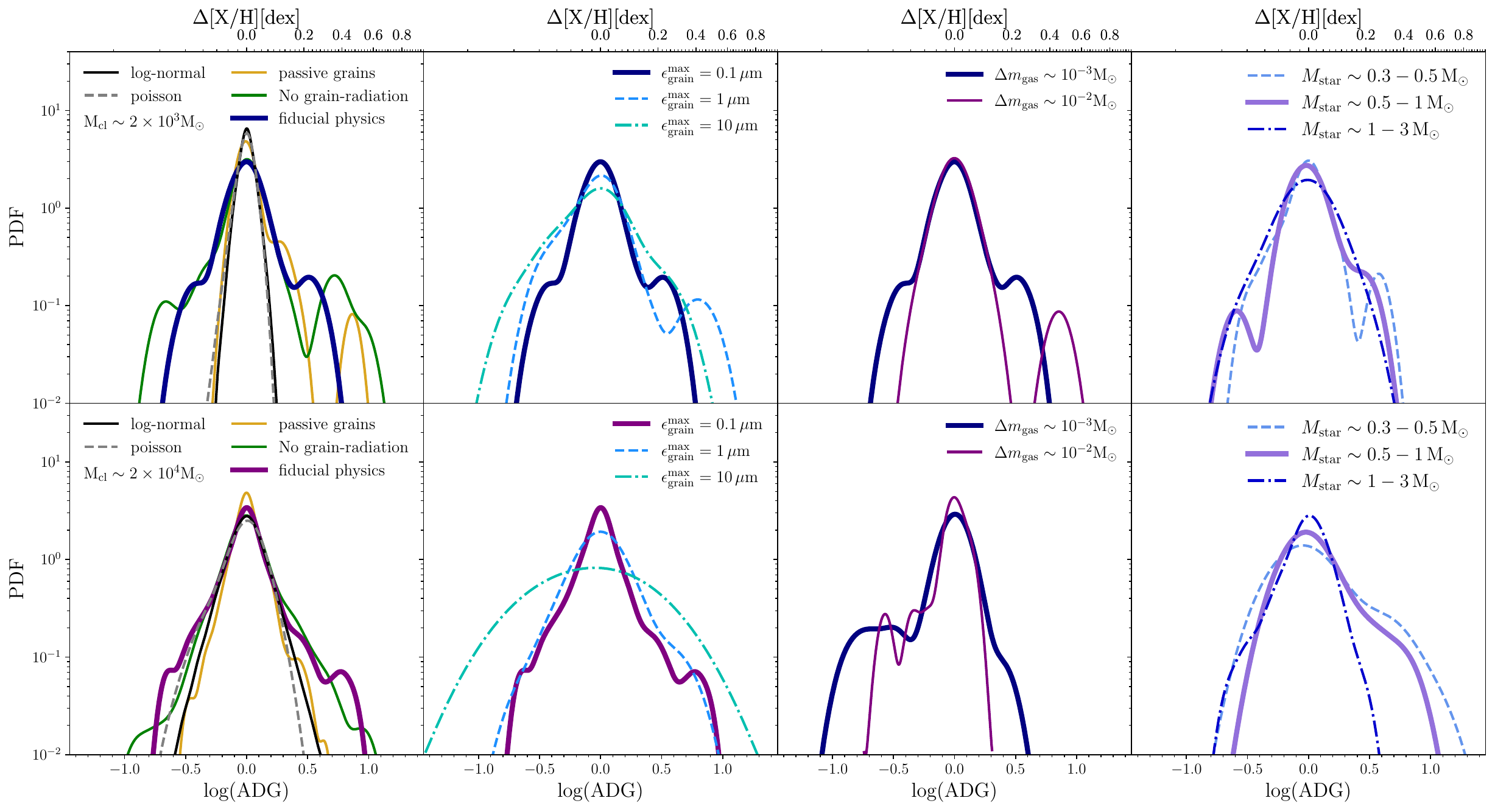}
    \caption{Probability density functions (PDFs) showing the deviation of accreted dust-to-gas ratio (ADG) and equivalent surface abundance variations in refractory elements $\rm \Delta [X/H]$ from the median for all stars of similar mass that form in a cloud of mass $M_{\rm cl} \sim 2\times 10^3 \rm M_\odot$ (top) and $M_{\rm cl} \sim 2\times 10^4 \rm M_\odot$ (bottom) simulations. From left to right, we examine the impact of different physics runs, grain sizes, resolutions, and stellar mass on these distributions. Comparisons are made in the leftmost plots with log-normal and Poisson distributions, both centered at $\rm \Delta [X/H]= 0$ with a standard deviation calculated as $\delta\sqrt{\dustgas M_{\rm star}/ \Delta m_{\rm dust}}$, where $\delta \sim 8$ is fitted to encompass the core width in simulations. Passive grain scenarios (dust only feels drag forces) yield narrower distributions compared to full physics and no radiation pressure cases, where additional physics drives dust-gas separation. Distributions broaden with larger grain sizes due to greater decoupling from gas dynamics. High-resolution runs show similar core widths to low-resolution runs but have more extended tails, likely due to better-resolved fluctuations. In high-resolution small cloud simulations, stellar mass ranges show similar widths, while in lower-resolution, higher-mass clouds, the width is dominated by less massive stars. These findings highlight the role of dust dynamics in driving variation in the surface abundances of refractory elements. }
    \label{fig:stack}
\end{figure*}

\begin{figure}
    \centering
    \includegraphics[width=1\linewidth]{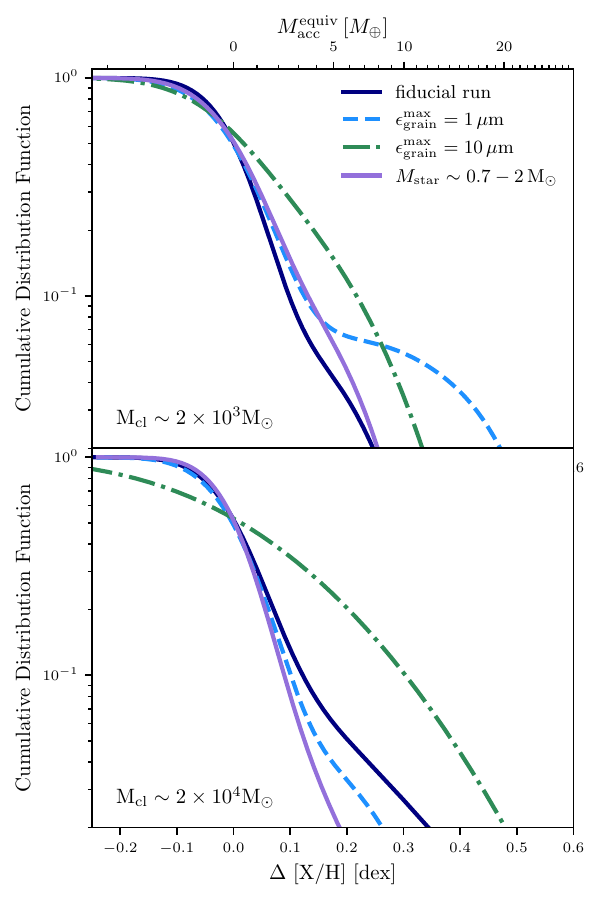}
    \caption{The cumulative distribution function of surface abundance variations in refractory elements \(\Delta \rm [X/H]\) and the equivalent mass needed to produce the same variation, corresponding to accretion of \(M_{\rm acc}^{\rm equiv}\) for a 1 \(\rm M_\odot\) star. Results from simulations with cloud sizes \(M_{\rm cl} \sim 2\times 10^3 \rm M_\odot\) (top panel) and \(M_{\rm cl} \sim 2\times 10^4 \rm M_\odot\) (bottom panel) show that 10-30\% of stars exhibit around a 0.1 dex variation in \(\Delta \rm [X/H]\), equivalent to accreting 2-5 \(\rm M_\oplus\) planets, consistent with \citep{liu2024least, spina2021chemical}. Comparing stars in the 0.7 to 2 \(\rm M_\odot\) range with those in the 0.1 to 10 \(\rm M_\odot\), typically covered in the broader analysis, reveals minor differences in simulations with smaller clouds at finer resolution. However, larger clouds at coarser resolution show deviations below 20\%, mainly due to variations in low-mass stars. Additionally, simulations with a maximum grain size of \(\grainsizemax = 1 \, \mu\text{m}\) yield results similar to those with the fiducial \(\grainsizemax = 0.1 \, \mu\text{m}\), with deviations occurring at rates below 10\%. In contrast, simulations with \(\grainsizemax = 10 \, \mu\text{m}\) exhibit greater variation due to the reduced coupling of larger grains with the gas dynamics.Overall, we find that dust dynamics naturally produce observationally equivalent signal to planet ingestion.}
    \label{fig:mass}
\end{figure}

\begin{figure*}
    \centering
    \includegraphics[width=1\linewidth]{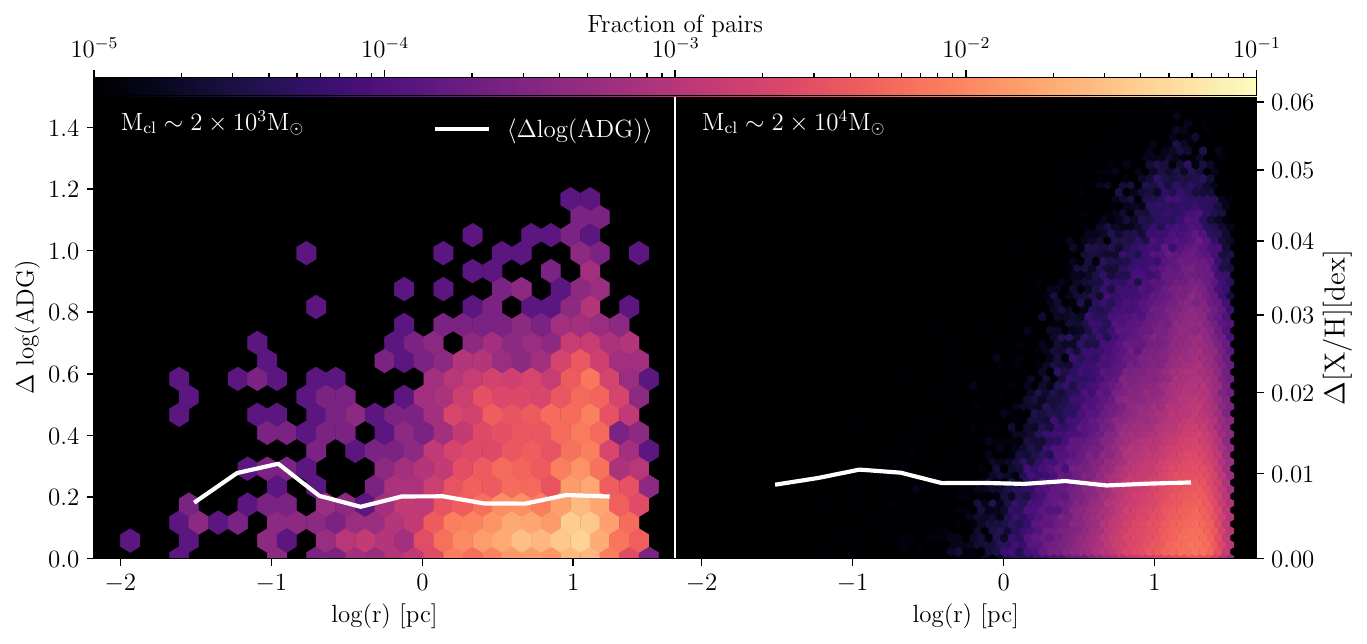}
    \caption{The 2D histogram depicts the bivariate distributions of the accreted dust-to-gas (ADG) ratio, the equivalent surface abundance variations in refractory elements $\rm \Delta [X/H]$, and stellar separation ($r$) for all pairs of stars (bound and unbound) after orbital relaxation in the final snapshot in the simulations. On the left, we present the distribution for a cloud of size $M_{\rm cl} \sim 2\times 10^3 \rm M_\odot$, and on the right, for a cloud of size $M_{\rm cl} \sim 2\times 10^4 \rm M_\odot$. A median line illustrates the deviation as a function of separation bins. The data reveal fluctuations spanning up to 1 dex. The absence of a pronounced trend with separation implies that the underlying physics governing these variations operates on scales where stellar separation exerts minimal influence.}
    \label{fig:r}
\end{figure*}

Figure \ref{fig:stack} illustrates the probability density function that represents the deviations from the logarithm of the accreted dust-to-gas ratio (ADG). The ADG, denoted as $\mu_{\rm acc}$, is defined as the ratio of the mass of accreted dust to the mass of accreted gas which can be approximated by the ratio of the mass of accreted dust to the stellar mass $M_{\star}$. This ratio is computed as:

\begin{align}
    \mu_{\rm acc} \approx \frac{\sum_{i=1}^{N_{\rm dust}} \Delta m_{\rm dust, i}}{\sum_{i=1}^{N_{\rm gas}} \Delta m_{\rm gas, i} + \sum_{i=1}^{N_{\rm dust}} \Delta m_{\rm dust, i}},
\end{align}

 The terms \(\Delta m_{\rm dust, i}\) and \(\Delta m_{\rm gas, i}\) denote the mass of individual dust and gas components, respectively, while \(N_{\rm dust}\) and \(N_{\rm gas}\) are the total numbers of the accreted dust and gas components.

For each bin of stellar mass, the deviation of the ADG from the median ADG value for stars of similar mass is calculated. This deviation is expressed in logarithmic terms as \(\mathrm{log} \left(\mu_{\rm acc}/\langle \mu_{\rm acc}\rangle \right)\), where \(\langle \mu_{\rm acc} \rangle\) represents the median ADG for stars within the same mass range. 

To analyze how the dust-to-gas ratio variations from the median values translate to changes in metallicity $\Delta \rm [X/H]$, we use the relation 
\begin{align}
    \Delta \rm [X/H] \equiv \rm log ( f_{\rm m} \mu_{\rm acc} / \langle \mu_{\rm acc} \rangle - (1- f_{\rm m})).
\end{align}

Here, $f_{\rm m} \sim 0.3$ represents the fraction of refractory metals depleted onto dust grains \citep{jenkins2009unified}. This calculation assumes homogeneous mixing of the total dust mass accreted by the star throughout its entire mass. Our analysis focuses on stars with masses between $0.3-10 \rm M_\odot$ in clouds with masses $M_{\rm cl} \sim 2 \times 10^{3} \rm M_{\odot}$ (top) and $M_{\rm cl} \sim 2 \times 10^{4} \rm M_{\odot}$ (bottom), ensuring a well-sampled population for statistical robustness.

Given the discretization of gas and dust by resolution, Poisson fluctuations in ADG and $\Delta \rm [X/H]$ are expected, scaling with $\sqrt{N_{\rm dust}}$, where $N_{\rm dust}$ is the number of accreted dust particles. To evaluate these fluctuations, we compared them with theoretical Poisson and log-normal distributions, both centered around $\Delta \rm log(ADG) = \Delta [X/H] = 0$, suggesting no deviations from the median. We found that the typical Poisson and log-normal error predicted from our resolution sampling is much narrower than the data. To ensure a conservative analysis, we fitted these distributions to cover the entire core of our data distribution, corresponding to a standard deviation of $\delta \sqrt{\frac{\dustgas M_{\rm star}}{\Delta m_{\rm dust}}} \sim \delta \sqrt{N_{\rm dust}}$, where the dust mass element resolution is $\Delta m_{\rm dust} = \Delta m_{\rm gas}/4$, and $\delta \sim 8$, equivalent to reducing the sampling resolution by a factor of 64.

We discuss each column in the figure in separate sections: fiducial results in \S\ref{sec:fid}, dust physics in \S\ref{sec: dustphys}, grain size distributions in \S\ref{sec:grainsize}, and numerical resolution and stellar mass ranges in \S\ref{sec:resol}.

\subsection{Fiducial findings}
\label{sec:fid}
We find that the distributions are described by a narrow peak and broad tails. Despite our very conservative approach, our data tails significantly exceed model predictions, suggesting that observed deviations are physical, not numerical. This is in agreement with findings from \citet{lee2017dynamics} and \citet{ moseley:2018.acoustic.rdi.sims}, which demonstrated significant dust-to-gas fluctuations in simulations of dust-gas dynamics in similar environments.

To compare our statistics with observations, Figure \ref{fig:mass} shows the cumulative distribution of $\Delta \rm [X/H]$ for stellar populations. We also compared our results with predicted planetary mass accretion if fluctuations were due to planetary ingestion, calculated as $M_{\rm acc}^{\rm equiv} \equiv \dustgas f_{\rm m} f_{\rm cz} M_{\rm sink}$ for a 1 $M_{\rm sink} \sim 1 \rm M_\odot$ star, assuming a convective zone fraction of $f_{\rm cz} \sim 0.01$. We found that in our fiducial simulations  $\sim$10-30\% of stars exhibit $\sim 0.1$ variation in $\Delta \rm [X/H]$. Under our assumptions, this variation translates to an accretion of $\sim 2-5 \rm M_\oplus$ of planetary material within the convective zone. These findings align closely with the statistics reported in previous observational studies \citep{liu2024least, spina2021chemical}.

Our qualitative findings remain robust; however, their quantitative aspects depend on parameters which will be examined in the succeeding sections.

\subsection{Influence of physical processes}
\label{sec: dustphys}
First, we consider the influence of physical processes on the variability of the ADG ratio. In the leftmost panel of Figure \ref{fig:stack}, we present a comparison among different physics runs. These include our full fiducial physics run, where grains experience radiative, magnetic, and drag forces with back-reaction on the gas. We compare it with a run excluding radiation pressure forces on grains (No grain-radiation), and another simulating passive grains experiencing only drag forces.

Our findings reveal that the passive grain run exhibits a narrower core-width compared to both the full physics and no radiation-dust coupling runs, which show comparable broadening. This observation aligns with theoretical expectations, as the passive grain simulations lack the physical mechanisms that drive dust-gas separation \citep{hopkins2018ubiquitous, hopkins2020simulating}. Consequently, this naturally results in a narrower variation in stellar abundances.

The similarity in distributions between the full physics and no radiation-dust coupling runs is unsurprising. As discussed in \citet{soliman2024dust}, radiation pressure primarily acts to systematically reduce the ADG ratio for $M_{\rm sink} \gtrsim 2 \rm M_\odot$ due to the stronger radiation field associated with more massive stars. Thus, any variations stemming from this effect manifest as offsets among stars of different masses. However, since our primary focus in this analysis is to emphasize the spread of the distribution rather than systematic offsets, any mass-dependent systematic trends are subtracted.

Furthermore, in the full-physics runs, while the core of the distribution could be matched to Poisson or log-normal noise by inflating sampling errors by a factor of 8, the tails of the distribution notably deviate from Gaussian or Poisson behaviors. Therefore, we conclude that the variation in core width, coupled with even greater variation in the tails of the distribution as more detailed dust physics are accounted for, suggests a physical origin rather than purely statistical effects.

\subsection{Impact of Grain Size Variation}
\label{sec:grainsize}
In addition, we investigate how different grain size distributions influence the distribution spread. In the scenario of infinitesimally small grains, we expect perfect dynamic coupling with the gas, resulting in a delta function around $\Delta \rm [X/H]= 0$. However, as grains increase in size, their coupling to the gas dynamics/gas accretion weakens, which could lead to larger variations in ADG ratios.

In the second column of Figure \ref{fig:stack} and \ref{fig:mass}, we present the probability density function and cumulative distribution respectively, for clouds with dust grain sizes ranging from $\grainsizemax = 0.1\rm \mu m$ to $\grainsizemax = 10\rm \mu m$. As anticipated, the distribution broadens with larger grain sizes, with $\sim 0.2$ dex more variation in $\Delta \rm [X/H]$ when $\grainsizemax$ increases to $10 \ \mu m$. For grain sizes of $0.1 \ \mu m$ and $1 \ \mu m$, the runs overlap and show no broadening trend, possibly due to resolution limits or similar dust-to-gas fluctuations at these sizes. A more pronounced difference is observed at $10 \rm \mu m$.

This broadening highlights that the physical properties of dust grains drive distribution changes, providing further evidence for a physical underpinning of the distribution's broadening, likely driven by dust dynamics. However, larger grain sizes correlate with fewer stars formed in the simulation, as discussed in \citet{soliman2024thermodynamics}, leading to less sampling of the tails for larger grains.

\subsection{Dependence on numerical effects and resolution}
\label{sec:resol}

To assess the robustness of our findings, we examine the effects of different resolutions on the distributions, depicted in the third panel from the left in Figure \ref{fig:stack}, considering resolutions of $10^{-3} \rm M_\odot$ and $10^{-2} \rm M_\odot$. If the fluctuations were purely numerical artifacts, we would expect the distribution widths to increase by a factor of $\sqrt{10}$ with the coarser resolution. However, we do not observe such broadening. Both high-resolution and low-resolution simulations yield similar core distributions, and extended tails persist even with a tenfold resolution increase.

We further validate this by analyzing the 25-75th percentile (core) and the 5-95th percentile (tails) intervals for each resolution. The 25-75th percentile remains constant or increases with higher resolution, while the 5-95th percentile width increases by approximately 0.3-0.4 dex with resolution refinement. This indicates that the fluctuations are not due to poor resolution but rather are intrinsic to the physical processes driving dust clumping and dust-gas separation, such as resonant drag instabilities, which can cause significant dust-gas separation and clumping across various scales \citep[e.g.][]{hopkins2022dust}.

Given that we are likely not resolving all small-scale structures and clumping, it is expected that the tails of the distribution will expand as more dust-rich structures are captured. Numerical convergence might not be fully achieved in our simulations due to the complexity of the physics involved. However, the smallest relevant scales, related to sonic turbulence in dense star-forming regions \citep{arzoumanian2011characterizing, roman2011turbulence, federrath2016universality, arzoumanian2018molecular}, are within our resolution limits.

Furthermore, the distribution’s narrow core and extended tails show that most stars have minor deviations from the median ADG ratio. However, the stars in the tails, with substantial abundance variations, may resemble a secondary population stars that accumulated surface abundance variations due to discrete events such as planet ingestion. 

In the right-most panel in Figure \ref{fig:stack}, we show the distributions for stars for $M_{\rm star} \sim 0.3-0.5 \rm M_\odot$,  $M_{\rm star} \sim 0.5-1 \rm M_\odot$, $M_{\rm star} \sim 1-3 \rm M_\odot$, each separately. We find that in each stellar mass, the distribution is broader than the log-normal and/or poisson distribution even with $\delta \sim 8$. Numerically, different stellar masses test resolution effects. If variations were due to numerical sampling, distribution widths would scale with $1/\sqrt{M_{\rm star}}$. However, widths do not consistently follow this scaling across different stellar masses. Despite varying resolutions, the distributions exhibit similar widths across different stellar masses. This suggests that the observed variations in distribution widths are unlikely to be purely numerical sampling effects.

\subsection{Dependence on stellar seperation}
\label{sec:sep}
In Figure \ref{fig:r} we show the variation in the ADG ratio between each pair of stars in the simulation after orbits have relaxed and the equivalent $\Delta \rm [X/H]$ as a function of stellar separation $r$. Interestingly, the data indicates a weak or negligible trend between ADG ratio variations and stellar separation. This suggests that the processes influencing these variations operate on relatively small spatial scales. Consequently, the physical mechanisms driving dust-gas separation and subsequent abundance variations are not significantly affected by the distances between stars within the same cluster. 

\subsection{Caveats}

This study represents an initial exploration and plausibility study of the impact of dust-gas dynamics on stellar refractory element abundance variations. While our findings align broadly with observed phenomena, several caveats and avenues for improvement should be considered:

\begin{itemize}

\item {\bf{Simplified stellar models}}: Our analysis employs a highly simplified stellar model, assuming uniform mixing of accreted dust throughout the star's mass without detailed simulation of internal stellar structures. To estimate the mass of planetary material in the convective layer, we assume that the convective layer constitutes 1\% of the stellar mass for Sun-like stars. Although mixing rates and convective layer thickness vary and evolve over time, this approach provides a first-order approximation linking preferential dust accretion to surface abundance variations.

\item {\bf{Simplified dust chemistry and evolution}}: We do not account for complex dust chemistry, variations in dust-to-metal ratios, species depletion. Our model assumes a constant metal fraction of $f_{\rm m} \sim 0.3$ on grains, though this can vary roughly from 0.2 to 0.5 \citep{jenkins2009unified}. Nevertheless, we conservatively estimate on the lower bound due to the depletion of metals onto polycyclic aromatic hydrocarbons (PAHs) and nanometer-sized grains. These components, owing to their elevated number abundance and large surface area-to-mass ratio, are well-coupled to gas dynamics and thus likely do not contribute to dust-gas segregation effects. Additionally, although our current simulations do not incorporate processes like dust growth, coagulation, and shattering, which play an  important role in setting the dust-to-metal ratio and setting the grain size distribution \citep{hirashita2011effects, hirashita2019remodelling, relano2020evolution}, we aim to explore their effects in future studies. 

\item {\bf{Resolution Scale}}: Our simulations do not resolve fluctuations occurring on small scales, such as fluctuations within a single core or disk, and those within accretion disks around individual stars. These unresolved scales could affect trends among binary stars, particularly among short-period binaries formed from common disk fragmentation. Additionally, our simulations do not resolve the formation of planetesimals or planets, which could sequester dust mass from the disk and prevent it from being accreted by the star.
\end{itemize}

Future efforts will focus on refining these models and addressing these limitations o provide a more comprehensive understanding of the processes involved.

\section{Conclusions}
\label{sec:conc}

In this study, we explored fluctuations in dust-to-gas ratios near stars as a source of surface abundance variations of refractory elements ($\Delta \mathrm{[X/H]}$) in stellar clusters using detailed star formation simulations. These simulations varied cloud masses, resolutions, and initial grain size distributions, incorporating comprehensive dust-gas dynamics. Our key findings include:

\begin{itemize}
    \item Our simulations predict that $\sim 10-30\%$ of stars show $\Delta \rm [X/H] \sim 0.1$ dex variation, equivalent to the accretion of a $2-5 \mathrm{M_\oplus}$ planetary object into the convective layer of Sun-like stars. 
    
    \item The $\Delta \rm [X/H]$ distribution features a narrow central peak with extended tails, which would give rise to two distinct populations: stars with standard abundance patterns and those with enhanced abundance patterns.
        
    \item Resolution comparisons show the extended tails of these distributions are robust and significantly different from log-normal or Poisson distributions, suggesting the abundance variations are not purely statistical.

    \item Simulations with full dust physics yield broader distributions compared to those with limited dust physics, underscoring the critical role of dust dynamics in shaping the $\Delta \rm [X/H]$ distribution.
    
    \item Larger grain sizes correlate with broader distributions, emphasizing the impact of dust grain properties on abundance variations.

    \item No significant correlation is found between abundance deviations and stellar separations down to 0.01 pc.

\end{itemize}

However, future investigation is needed to refine our understanding of stellar chemical enrichment mechanisms, as our approach involved simplified stellar evolution models, dust chemistry, and resolution limitations.

We note that our model does not dispute the occurrence of planetary engulfment but suggests that similar fluctuations in refractory surface element abundances can arise from variations in dust-to-gas ratios during star formation. Distinct from engulfment, this alternative mechanism predicts that abundance variations would persist throughout a star's lifetime, including earlier stages with thicker convective zones, and could drive both reductions and enhancements in refractory surface abundances.

 In summary, our study highlights the importance of considering dust-gas fluctuations as a  source of chemical enrichment of stars. Further research is required to quantify the relative contributions of different enrichment processes and to refine our understanding of chemical homogeneity within stellar associations.

\datastatement{The data supporting this article are available on reasonable request to the corresponding author.}

\begin{acknowledgments}
We thank Caleb R. Choban for his valuable feedback on the clarity and organization of this work. We also thank Carlos Saffe for his insightful suggestions and observations, which helped us refine our research. Support for for NS and PFH was provided by NSF Research Grants 1911233, 20009234, 2108318, NSF CAREER grant 1455342, NASA grants 80NSSC18K0562, HST-AR-15800. Numerical calculations were run on the TACC compute cluster ``Frontera,'' allocations AST21010 and AST20016 supported by the NSF and TACC, and NASA HEC SMD-16-7592.
\end{acknowledgments}

\software{\href{https://matplotlib.org/}{\fontfamily{cmtt} \selectfont matplotlib} \citep{Hunter:2007}, \href{https://numpy.org/}{\fontfamily{cmtt} \selectfont NumPy} \citep{harris2020array}, \href{https://scipy.org/}{ \fontfamily{cmtt} \selectfont SciPy} \citep{2020SciPy-NMeth}.}

\bibliography{sample631}{}
\bibliographystyle{aasjournal}
\end{document}